\documentclass[twocolumn,preprintnumbers,superscriptaddress,endnote,nofootinbib,aps,prd,floatfix]{revtex4}

\usepackage{footmisc}
\usepackage{subfigure}
\usepackage{amsmath}
\usepackage[utf8]{inputenc}
\usepackage{graphicx}
\usepackage{hyperref}
\hypersetup{colorlinks=true, citecolor=blue, urlcolor=blue, linkcolor=blue}

\preprint{IPPP/18/61} 

\begin{document}

\title{Machine Learning Uncertainties with Adversarial Neural Networks}

\begin{abstract}
Machine Learning is a powerful tool to reveal and exploit correlations in a multi-dimensional parameter space. Making predictions from such correlations is a highly non-trivial task, in particular when the details of the underlying dynamics of a theoretical model are not fully understood. Using adversarial networks, we include a priori known sources of systematic and theoretical uncertainties during the training. This paves the way to a more reliable event classification on an event-by-event basis, as well as novel approaches to perform parameter fits of particle physics data. We demonstrate the benefits of the method explicitly in an example considering effective field theory extensions of Higgs boson production in association with jets.
\end{abstract}

\author{Christoph Englert} \email{christoph.englert@glasgow.ac.uk}
\affiliation{SUPA, School of Physics and Astronomy, University of
  Glasgow,\\Glasgow, G12 8QQ, U.K.\\[0.2cm]}

\author{Peter Galler} \email{peter.galler@glasgow.ac.uk}
\affiliation{SUPA, School of Physics and Astronomy, University of
  Glasgow,\\Glasgow, G12 8QQ, U.K.\\[0.2cm]}

\author{Philip Harris} \email{philip.coleman.harris@cern.ch}
\affiliation{Massachusetts Institute of Technology, Department of Physics, Cambridge MA 02139, U.S.A.\\[0.2cm]}

\author{Michael Spannowsky} \email{michael.spannowsky@durham.ac.uk}
\affiliation{Institute for Particle Physics Phenomenology, Department
  of Physics,\\Durham University, Durham DH1 3LE, U.K.\\[0.2cm]}

\maketitle

\section{Introduction}
\label{sec:intro}
The application of multi-variate analysis (MVA) techniques and machine learning have a long-standing history in analyses in particle physics and beyond. In the context of particle physics, machine learning-based approaches are typically employed when the expected signal count is small compared to the expected background contribution, thereby challenging a more traditional cut-and-count analysis to reach sufficient discriminating power to separate signal from backgrounds. For instance, the recent observations of top quark-associated Higgs production by CMS~\cite{Sirunyan:2018hoz} and ATLAS~\cite{Aaboud:2018urx} heavily rely on multi-variate approaches. But machine learning has also been considered in different contexts. The power of MVAs in searches for new physics is that they adapt to correlations in particle final states in order to map out relations between theoretical input parameters (the Lagrangian) and the output, e.g. the physical final state given by a particular radiation profile observed in a detector~\cite{Komiske:2016rsd,Barnard:2016qma, Butter:2017cot,Cohen:2017exh,Chang:2017kvc,Pearkes:2017hku,Louppe:2017ipp,Kasieczka:2017nvn, deOliveira:2017pjk,Luo:2017ncs,Datta:2017lxt,Larkoski:2017jix,Shimmin:2017mfk,Metodiev:2017vrx,Roxlo:2018adx,Brehmer:2018kdj,Brehmer:2018eca,Collins:2018epr,Duarte:2018ite,Fraser:2018ieu,Komiske:2018oaa,Macaluso:2018tck,Andreassen:2018apy,deCastro:2018mgh,DAgnolo:2018cun,Brehmer:2018hga,Monk:2018zsb,Moore:2018lsr,DeSimone:2018efk}. 

\begin{figure*}[!t]
\includegraphics[height=6cm]{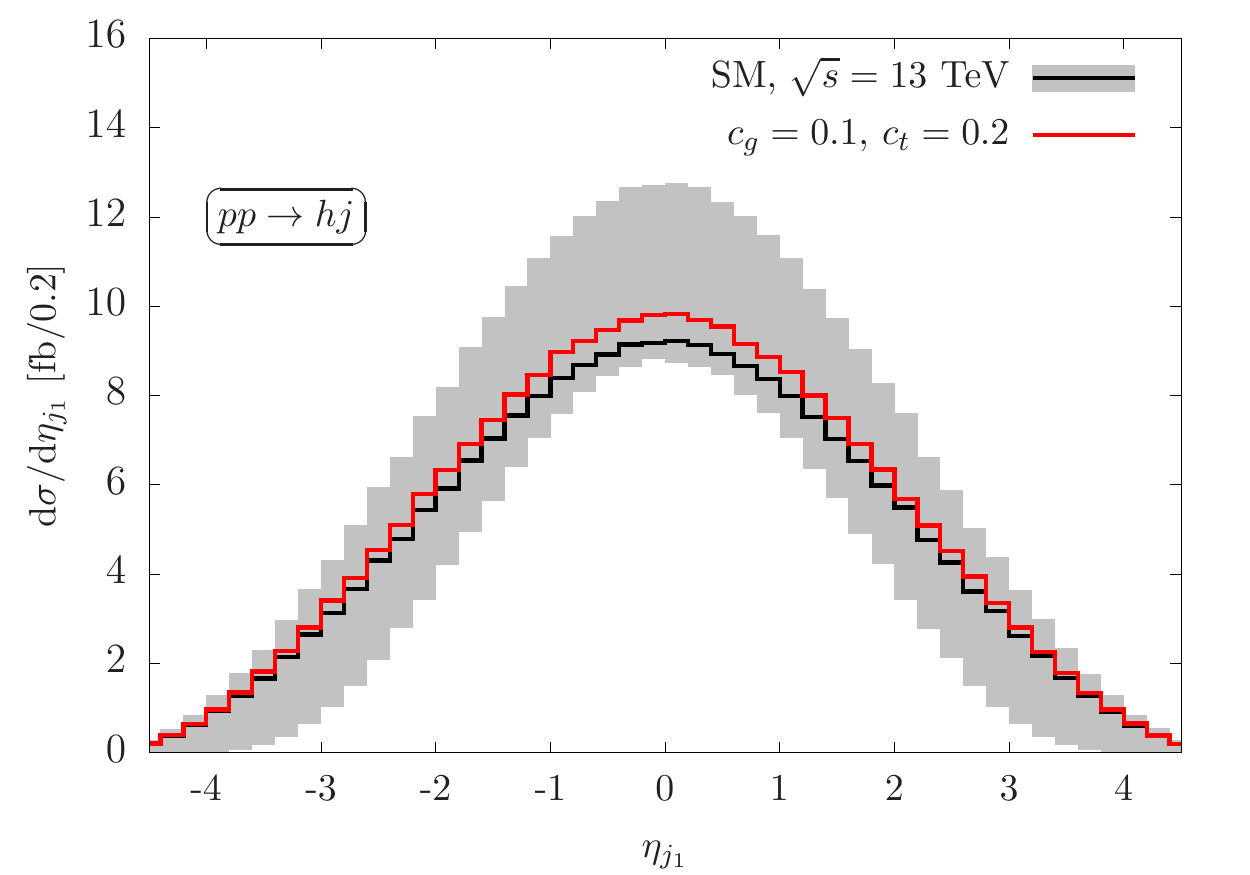}\hfill
\includegraphics[height=6cm]{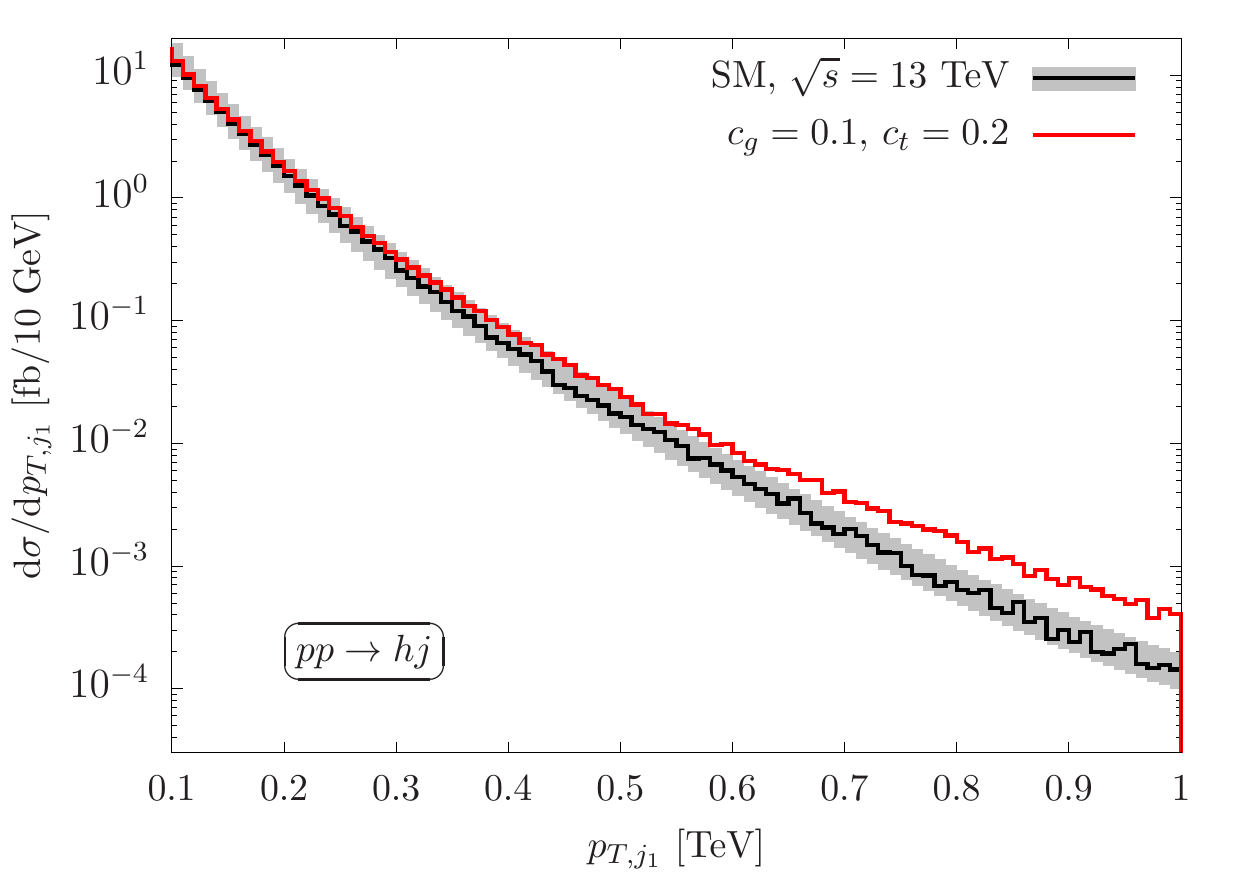}
\caption{\label{fig:hj} Predictions for Higgs+jet production for approximate cancellations of Wilson coefficient choices that can be resolved for large momenta. The uncertainty (grey band) is evaluated by factorisation and renormalisation scale variations ($\mu_0/2\leq \mu \leq 2\mu_0$) around the central scale $\mu_0=\sqrt{(p_h+p_j)^2}$. Modified branching ratios $h\to \tau\tau$ are included throughout.}
\end{figure*}

Machine learning approaches come into their own when there is insufficient knowledge of the dynamics that connect input and output, or in cases where there is no concrete model at all. This forms the basis of applications of machine learning approaches to stock trading and face or pattern recognition, where comparably effortless predictions need to be made on short timescales. This is qualitatively different for particle physics applications where the underlying Standard Model of Particle Physics (SM) is well-established. Connecting theoretical (not necessarily physical) input parameters with actual measurements is not only possible, but sets the baseline of the observed success of the SM over orders of magnitude. Of course, these strategies, which are supported by factorisation principles~\cite{Collins:1981tt,Collins:1985ue} at the price of associated uncertainties in perturbation theory, generalise to interactions beyond the SM. Therefore, the most adapted approach to classifying experimental observations (e.g. discriminating between signal and background) is using the theoretical model itself by employing its $S$-Matrix as an observable. This is known as the matrix-element method~\cite{Kondo:1988yd} and ATLAS and CMS have used these techniques in~Refs.~\cite{Khachatryan:2015ila,Aad:2015gra}. This approach can be extended to the full particle-level as discussed in~Refs.~\cite{Soper:2011cr,Soper:2012pb,Soper:2014rya, Englert:2015dlp}.

The downside of such methods is that they require extensive computational resources and quick event-by-event selection is not possible without further simplifying assumptions. These shortcomings motivate MVAs as interpolating tools whose sensitivity will be bounded by the sensitivity that could be achieved by a particle-level matrix element method.

Theoretical uncertainties are inherent to both the matrix element method as well as the multivariate techniques as the underlying Monte Carlo (MC) tool chain will be plagued by a range of largely unphysical parameter choices (e.g. renormalisation, factorisation and shower scales). MVAs need to be trained on MC output, at least for constraining models of new interactions or rare processes. Consequently, they inherit all MC-associated uncertainties. The MVA score will favour highly exclusive phase space region which are poorly understood perturbatively, enhancing the sensitivity to the underlying theoretical uncertainty. Data-driven methods might not be available in these very exclusive regions, and the price of a comparably large sensitivity is a reduced safety margin.

However, there are no well-defined models that can systematically estimate theoretical uncertainties. The impact of such effects is therefore estimated by the community's ad-hoc consensus on scale variations etc. This motivates MVAs as an ideal choice to decide on {\emph{how}} to propagate such unknowns to the final discriminant. This transcends the traditional envelope of kinematic observables or cross sections as the MVA will be equipped to ``see'' and extrapolate correlations of uncertainties and can decide on an event-by-event basis whether a particular configuration is sensitive to the question we might ask and whether the information we would like to draw from it can be trusted.

Such an approach provides unique opportunities to the extraction of unknown parameters. In particular, existing constraints from the LHC have left an impression that new physics could be heavy. This has motivated the use of effective field theory techniques for the hunt of new BSM interactions. The relevance of differential distributions in this context has been highlighted in~Refs.~\cite{Ellis:2014dva,Englert:2015hrx,Corbett:2015ksa,Englert:2017aqb} and the interplay of theoretical uncertainties in this context is extremely important. 

In this paper we extend existing machine learning techniques of treating systematic uncertainties using adversarial neural networks~\cite{Louppe:2016ylz} and propose a novel approach to include \textit{theoretical} uncertainties. In contrast to systematic uncertainties, which affect the kinematics on an event-by-event basis, theoretical uncertainties of the cross section are a property of the process at hand and affect the event sample as a whole. The ability to include all relevant uncertainties simultaneously not only allows for the evaluation of a neural network (NN) score in a much more controlled and meaningful way, but also paves the way to perform differential parameter fits on an event by event basis while fully including a measure of trust for the observed phase space region. We discuss this using the example of Higgs production in association with jets. However, our approach is applicable to a very wide range of scenarios where machine learning is used in  the presence of previously known theoretical and systematic uncertainties, e.g. signal vs background classification, particle identification/tagging and fitting of model parameters.

This paper is structured as follows: In Sec.~\ref{sec:eft}, we motivate Higgs+jets physics as BSM case where uncertainties are limiting factors in disentangling top-Yukawa modifications from gluon-Higgs contact interactions. In Sec.~\ref{sec:ann}, we  review the basics of the application of adversarial neural networks to controlling such uncertainties and highlight the power of this approach with a basic example, before we consider the full kinematics of Higgs production up to 2 jets in Sec.~\ref{sec:secapp}. We summarise and conclude in Sec.~\ref{sec:conc}.

\begin{figure*}[!t]
\includegraphics[height=6cm]{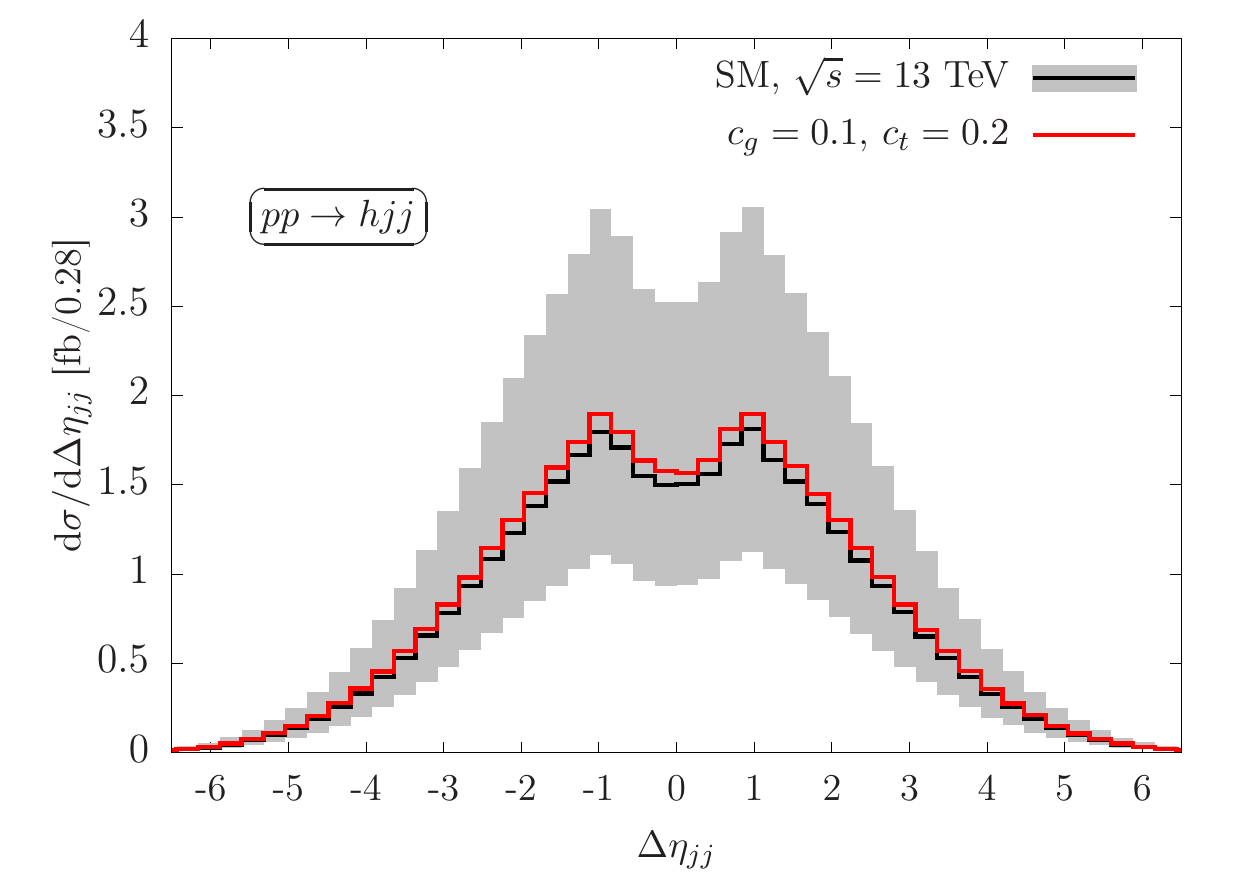}\hfill
\includegraphics[height=6cm]{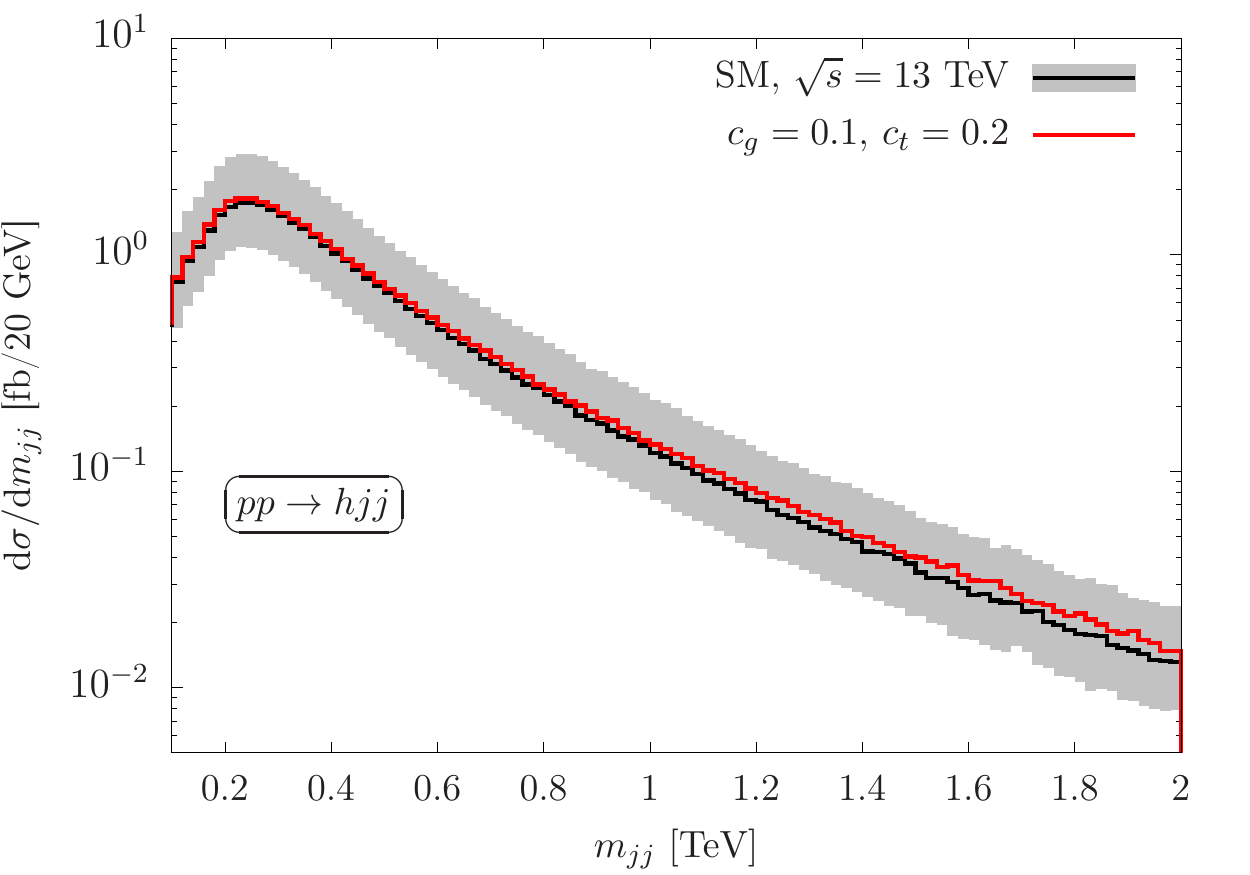}\\
\includegraphics[height=6cm]{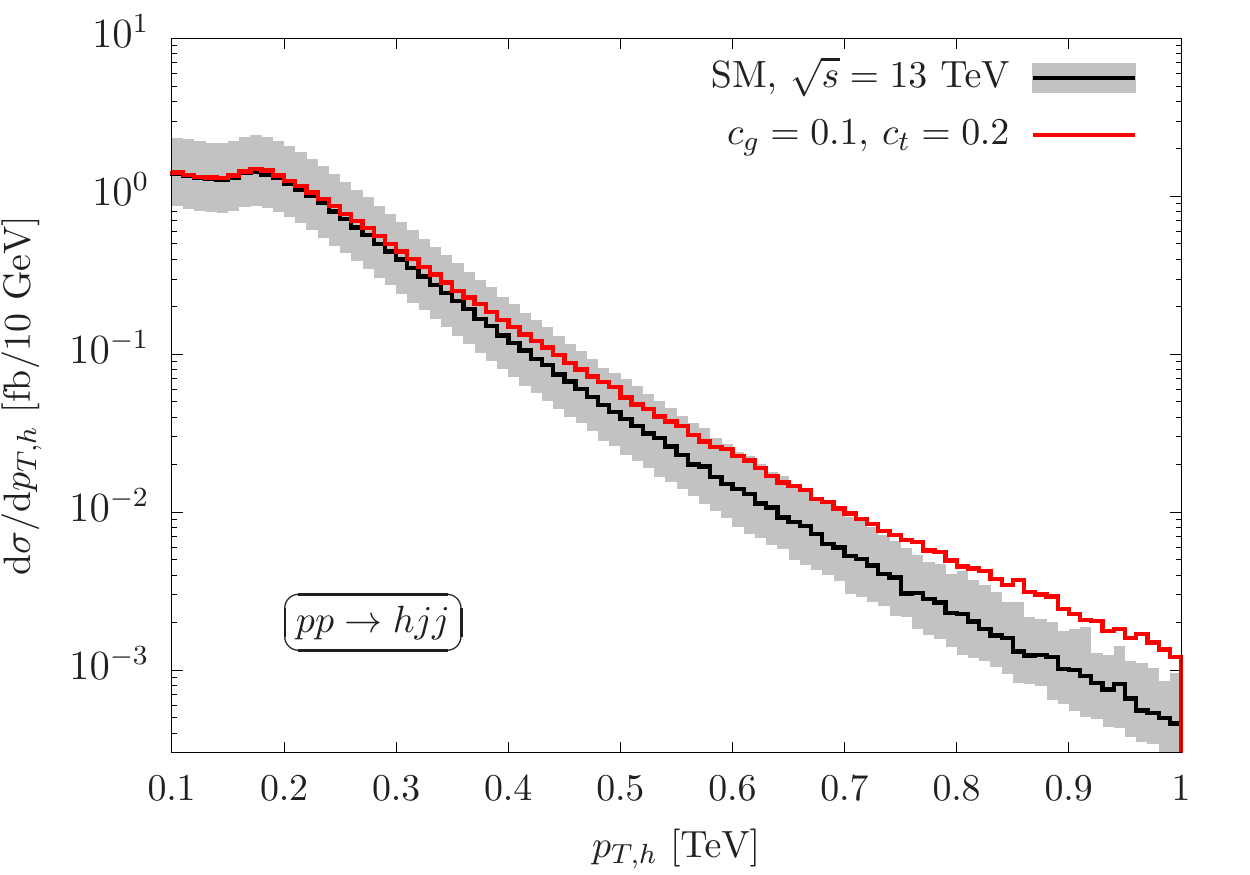}\hfill
\includegraphics[height=6cm]{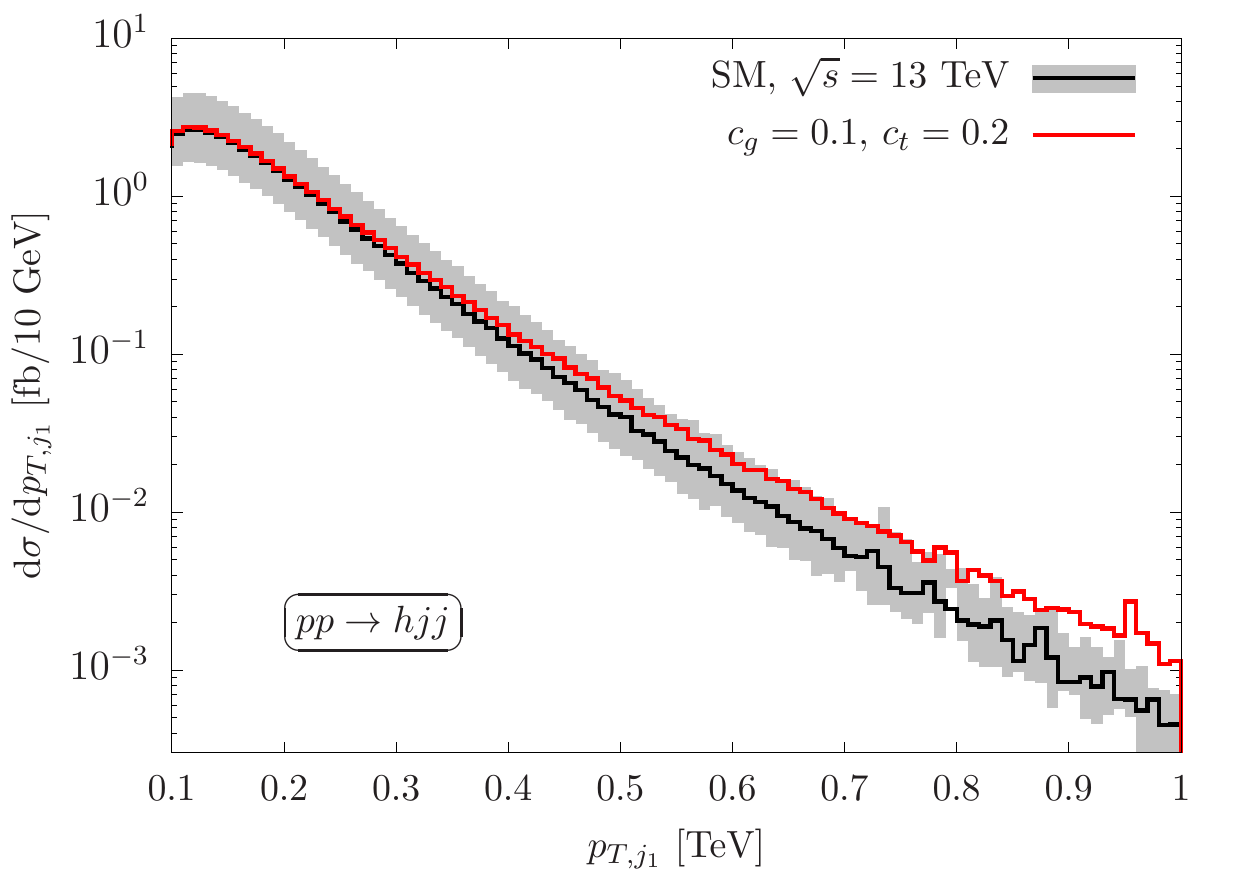}
\caption{\label{fig:hjj} Predictions for production in association with 2 jets for approximate cancellations of Wilson coefficient choices that can be resolved for large momenta. The uncertainty (grey band) is evaluated by factorisation and renormalisation scale variations ($\mu_0/2\leq \mu \leq 2\mu_0$) around the central scale $\mu_0= \sqrt{p_{T,j_1} p_{T,j_2}}$. Modified branching ratios $h\to \tau\tau$ are included throughout.}
\end{figure*}

\section{EFT measurements and differential distributions}
\label{sec:eft}
Extracting as much information as possible from energy-dependent observables is key to over-constraining the various parameters that need to be introduced if the low energy effects of new high-scale physics are treated generically~\cite{Ellis:2014dva,Englert:2015hrx,Corbett:2015ksa,Englert:2017aqb}. In particular, the high-$p_T$ regions of Higgs production can serve to break degeneracies of modified top quark-Higgs and effective gluon-Higgs interactions, which can be parameterised by
\begin{multline}
\label{eq:lag}
{\mathcal{L}}_{\text{d6}}=  c_g {\mathcal{O}}_g + c_t {\mathcal{O}}_t  = {c_g\, g_s^2 \over 16\pi^2 v} \, h\, G^{a\,\mu\nu} G^{a}_{\mu\nu} + c_t\, h\, \bar t t\,,
\end{multline}
where $G^{a}_{\mu\nu}$ denotes the gluon field strength tensor, and $h$ and $t$ the physical Higgs boson and top quark, respectively. The Wilson coefficient normalisations are chosen to make their numerical impact comparable (see below) and reflect the strongly-interacting light Higgs ansatz~\cite{Giudice:2007fh}, the additional factor of the strong coupling $g_s^2$ re-sums large logarithmic corrections from QCD at the dimension-6 level~\cite{Grojean:2013kd,Jenkins:2013zja,Englert:2014cva}. 

The top-Yukawa coupling modification at fixed top quark mass that is described by Eq.~\eqref{eq:lag} leads to a degeneracy with $c_g$ for momentum transfers below the top pair threshold. Concretely, low-energy theorems~\cite{Ellis:1975ap,Shifman:1979eb,Vainshtein:1980ea,Voloshin:1985tc,Kniehl:1995tn} induce interactions
\begin{equation}
{\mathcal{L}}_{\text{eff},t}= - {\sqrt{2}\over 3}{c_t \over y_t} {\mathcal{O}}_G + \dots \,, 
\end{equation}
where $y_t\simeq 1$ denotes the SM Yukawa coupling. This leads to an approximate blind direction \mbox{$\sim c_g-\sqrt{2}c_t/3$} of inclusive observables (such as cross sections), where the inclusive gluon fusion cross section becomes SM-like.

This degeneracy can be lifted in a global fit through subsidiary measurements of top quark-associated Higgs production, which is insensitive to the $ggh$ modifications~\cite{Englert:2015hrx}. Another promising avenue is to distinguish ${\mathcal{O}}_G$ from ${\mathcal{O}}_t$ at large momentum transfers~\cite{Banfi:2013yoa,Grojean:2013nya,Buschmann:2014sia,Buschmann:2014twa,Schlaffer:2014osa}, see Figs.~\ref{fig:hj} and \ref{fig:hjj}. The expected uncertainties in these particular phase space regions are non-negligible and are the obvious limiting factors of a coupling extraction from the theoretical side. 

Multiple hard jet emission can enhance the $c_g, c_t$ discrimination (see also~\cite{Duff:1991ad,Dreiner:1991xi,Dixon:1993xd,Krauss:2016ely} for related discussions). On the one hand, this comes at the price of an increased phase space suppression and a typically larger theoretical uncertainty. Additionally, the higher dimensionality of the phase space can give rise to new sensitive observables which are not necessarily directly aligned with standard kinematical distributions such as invariant mass and transverse momentum distributions. Adapting into these particular phase space regions can be achieved through boosted decision trees and other neural net techniques, which exploit multi-dimensional correlations to isolate particularly sensitive phase space regions. The downside of such an approach is that the associated uncertainties are hard to control, which can make such multivariate analyses highly sensitive to theoretical systematics. The case of Higgs production in association with multiple hard jets in the presence of ${\mathcal{O}}_g$, ${\mathcal{O}}_t$ modifications is particularly difficult and consequently provides a compelling physics case for the application of adversarial neural networks.

\subsection{Numerical setup}

In order to study the presence of ${\mathcal{O}}_g$ and ${\mathcal{O}}_t$ in Higgs production in association with hard jets we employ a modified version of {\sc{Vbfnlo}}~\cite{Campanario:2010mi,Arnold:2008rz,Baglio:2014uba} to perform the parton-level calculations presented in this work. Specifically, we focus on QCD-mediated Higgs production (gluon fusion) with one and two additional jets in the final state \cite{DelDuca:2001fn,DelDuca:2001eu,DelDuca:2001ad,DelDuca:2003ba,DelDuca:2006hk,Campbell:2006xx,Andersen:2010zx}. We pre-select events at the parton level in the central part of the detector with large cuts on the jet-transverse momentum distribution of 
\begin{equation}
\begin{split}
h+1~{\text{jet}}: &\quad 
p_{T,j}\geq 130~\text{GeV},~|\eta_j|<2.5 \\
h+2~{\text{jets}}: &\quad 
p_{T,j}\geq 150~\text{GeV},~|\eta_j|<4.5 
\end{split}
\end{equation}
to guarantee that these processes are well described by the associated hard matrix elements and that weak boson fusion and associated Higgs production can be controlled. For the chosen jet $p_T$ cut the weak contribution to $h$+2 jet production is around $1/5$. This contribution, which can be modified by other EFT operators is not discussed here and should be included in a more realistic EFT fit. Under these assumptions, the dominant Higgs coupling modifications to described Higgs production are parametrised by Eq.~\eqref{eq:lag}. We consider Higgs decays to tau leptons taking into account the branching ratio modifications induced by $c_g$ and $c_t$. We include $\tau$ tagging efficiencies independent of $c_g$, $c_t$ and phase space, but note that these are not major limiting factors at the LHC. In particular hadronic tau leptons are now under good control in Higgs final states, and di-tau efficiencies of around 50\% are possible at background rejection close to unity~\cite{Kreis:2015jjr,Cadamuro:2017slr,Dev:2017lde}. For computing significances for different choices of the Wilson coefficients $c_g$ and $c_t$ in Sec.~\ref{sec:secapp} we include a production reconstruction efficiency of 22\%~\cite{Englert:2015hrx} as well as a combined effective tau reconstruction efficiency of 43\%, which includes both leptonic and hadronic tau decay channels.

The theoretical uncertainties associated with the residual renormalisation ($\mu_R$) and factorisation ($\mu_F$) scale dependence of the observables are estimated by varying these scales around a central scale $\mu_0$
\begin{eqnarray}
\begin{array}{c}
\mu=\mu_R=\mu_F=\,\mu_0/2,\,\mu_0,\,2\mu_0\,,\\[0.3cm]
\mu_0 = \left\{\begin{array}{ll}
m_{hj}=\sqrt{(p_h+p_j)^2} & h+\text{jet}\\[0.1cm]
\sqrt{p_{T,j_1}p_{T,j_2}} & h+2~\text{jets}
\end{array}\right.,
\end{array}
\label{eq:scales}
\end{eqnarray}
where $m_{hj}$ is the invariant mass of $h$+jet and $p_{T,j_1}$ ($p_{T,j_2}$) is the transverse momentum of the (second) leading jet.

For this study we do not include a parton shower or detector simulation in the generation of $h+$jet and $h+2$~jets events because these effects are inconsequential to the method of including theoretical uncertainties using an adversarial neural network described in this work. The reason is that this method is based on supervised learning with Monte-Carlo events as input. Whether these events are evaluated at the parton, particle or detector level is not essential for the method to work. However, we expect parton shower and detector simulation to show some effect on the significances presented in Fig.~\ref{fig:ann} and defer the investigation of these effects to future studies.

\begin{figure*}[!t]
\includegraphics[width=0.48\textwidth]{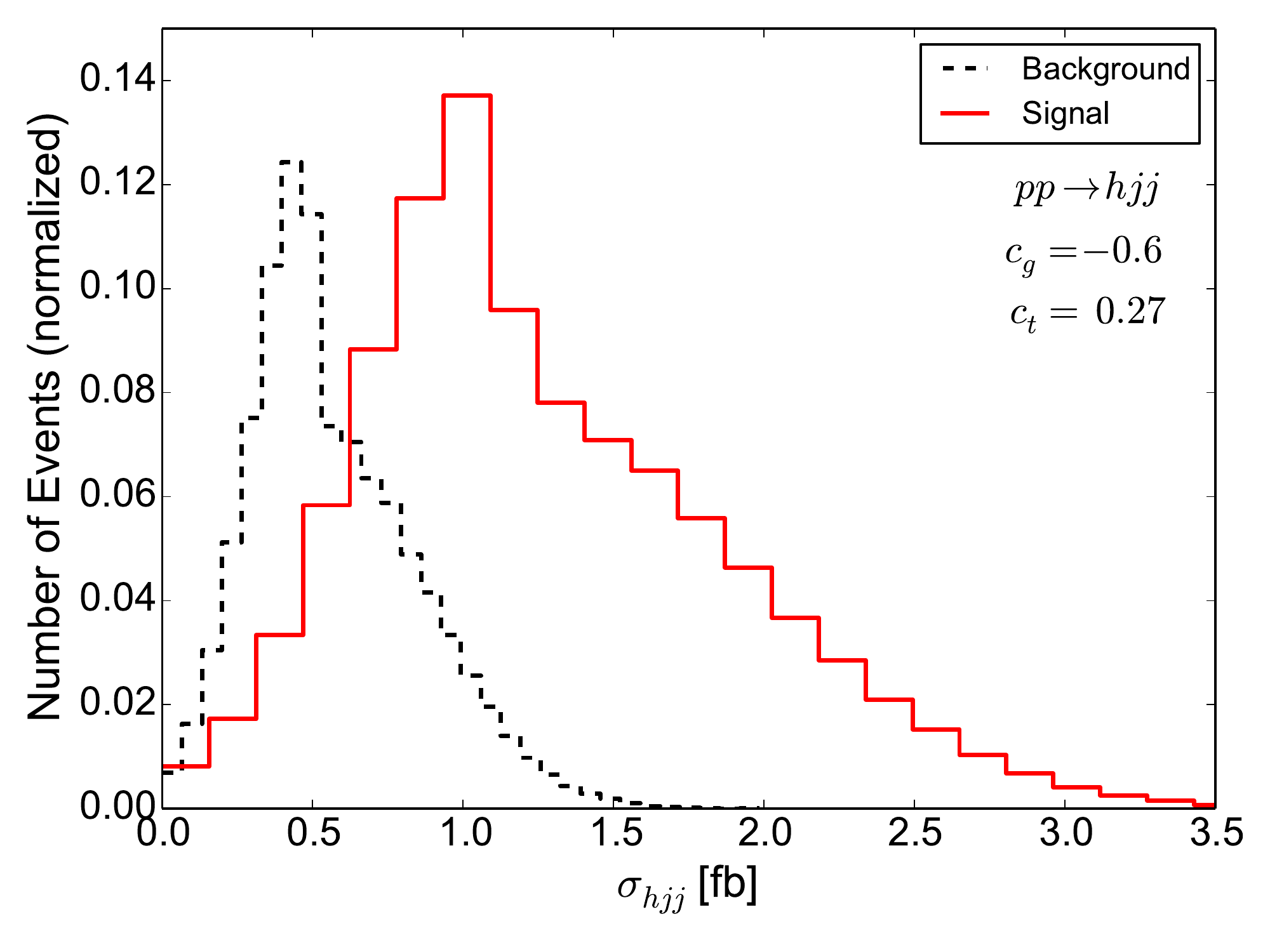}
\hfill
\includegraphics[width=0.48\textwidth]{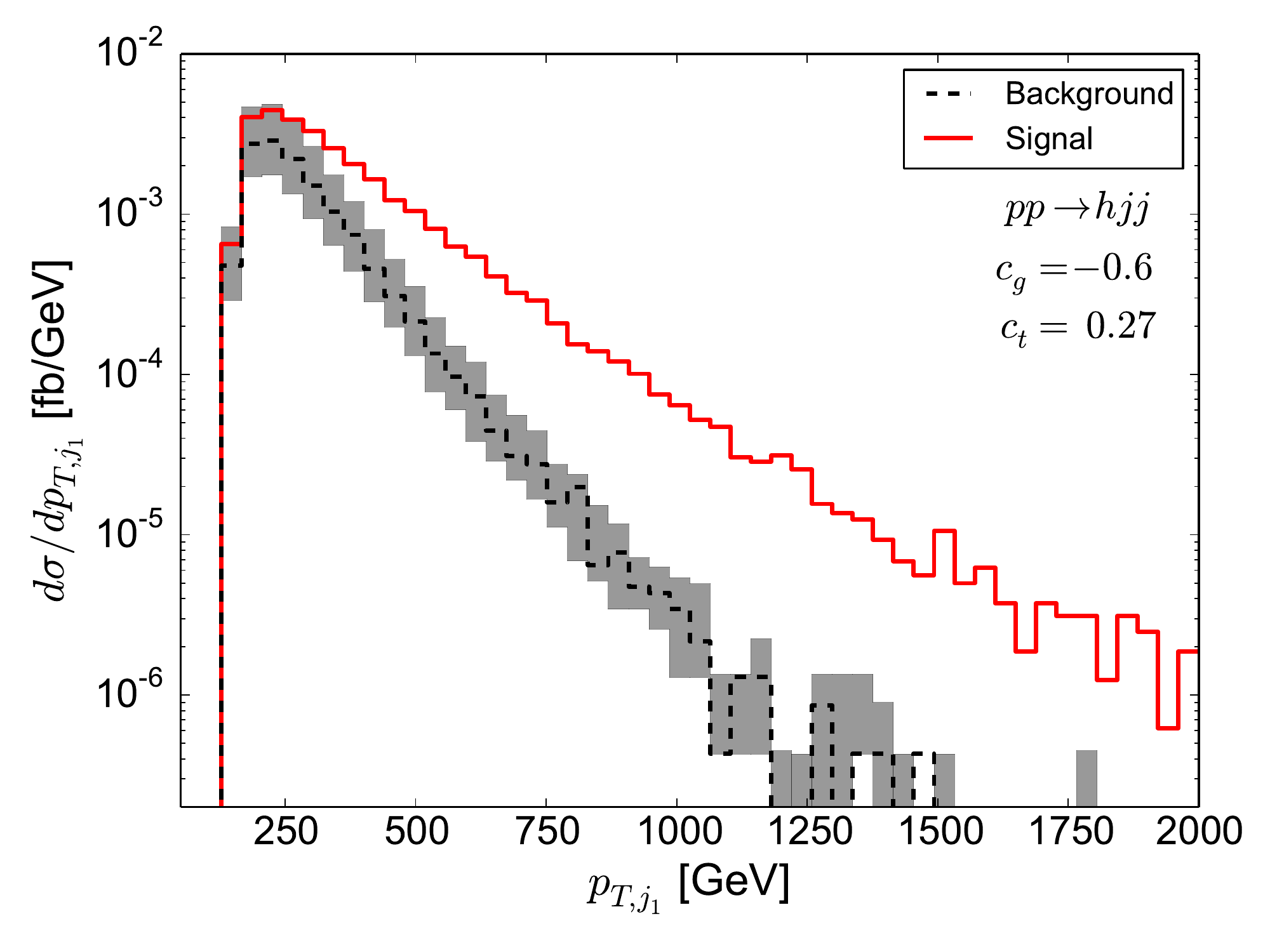}
\caption{\label{fig:nnexample}Cross section observable and jet-transverse momentum distribution in $h+2$~jets production for an operator choice $(c_g,c_t)=(-0.6,27)$. For further details see text.}
\end{figure*}

\section{Adversarial Neural Networks and Uncertainties}
\label{sec:ann}
\subsection{Learning uncertainties}
The concept of generative adversarial neural networks was first proposed in Ref.~\cite{Goodfellow:2014upx}. Its aim is to train a NN to generate data according to a given
(experimental) multi-dimensional distribution through a zero sum game. The setup consists of two NNs: a classifier and an adversary, which simultaneously use opposite training goals. The adversary learns to generate data samples according to the input distribution, while the classifier learns to distinguish generated from actual data. After the training of the setup when the NNs reach equilibrium, the classifier can only distinguish generated and real data by chance.

We make use of this approach by starting with a classifier that can distinguish between different input data variations according to the systematic uncertainties. The adversary on the other hand penalises this kind of discrimination via the loss function. The result of this adversarial training is a classifier that cannot distinguish between different input data variations and is therefore insensitive to the systematic uncertainties~\cite{Louppe:2016ylz}. More specifically, we can obtain a classifier into signal and background independent of underlying nuisance parameters such as theoretical uncertainties including the renormalisation and factorisation scale dependence. This is achieved by using the adversary to penalise the classifier whenever it becomes sensitive to the scale variation. The classifier thus avoids phase space regions that have a large discriminating power, but are plagued by theoretical uncertainties. This is the region relevant to disentangling different EFT contributions as discussed in Sec.~\ref{sec:eft}.

In total, such an adversarial neural network (ANN) is a numerical implementation of an optimisation problem (with respect to signal-background separation) with constraints (being independent of the scale) where the constraints are implemented via the loss function of the adversary and the associated Lagrange multiplier is a tunable hyper-parameter of the adversarial neural network.

Applying this to our physics problem, Monte Carlo runs with different scale settings can be used as input for the adversarial setup to discard phase space regions where discrimination also distinguishes the scale variations.
\begin{figure*}[!t]
\subfigure[]{\includegraphics[width=0.48\textwidth]{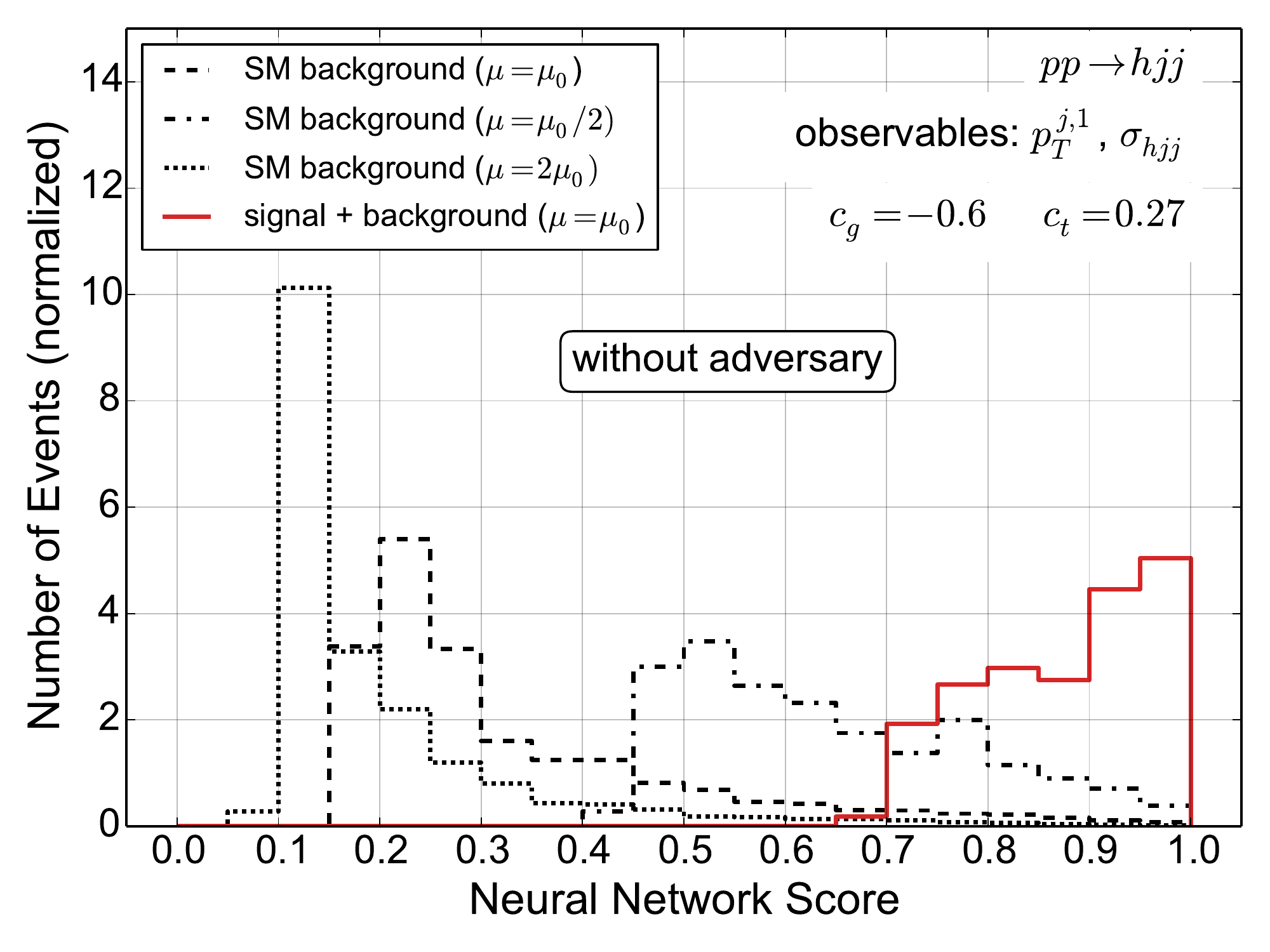}}
\hfill
\subfigure[]{\includegraphics[width=0.48\textwidth]{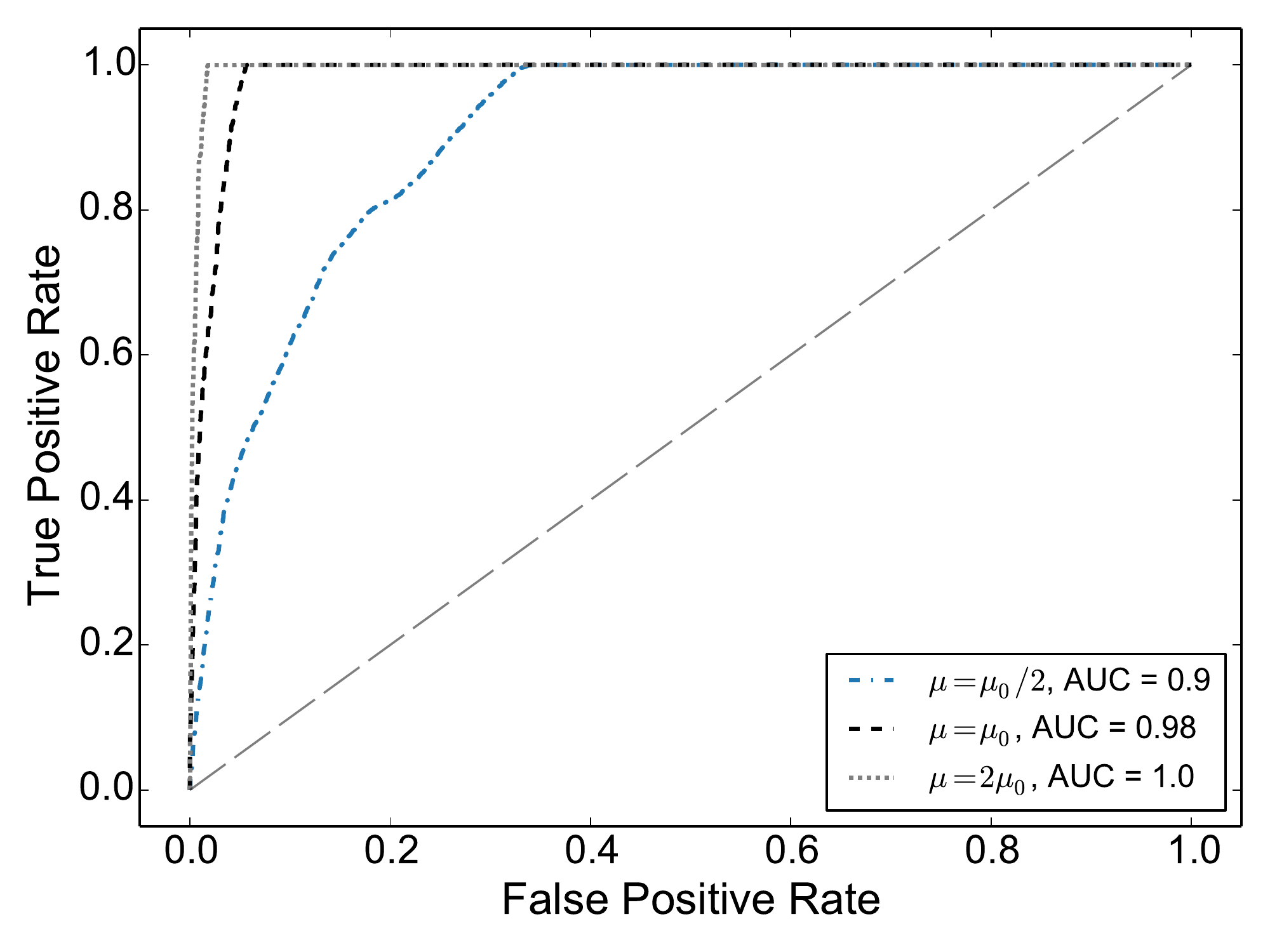}}
\caption{\label{fig:nnexamplescorenoadv} Distribution of NN scores (a) and associated ROC curve (b) for background-only and signal + background event samples. The classification has been performed using only the discriminator. If the area under curve (AUC) is larger than 0.5, discrimination is possible.}
\end{figure*}
\begin{figure*}[!t]
\subfigure[]{\includegraphics[width=0.48\textwidth]{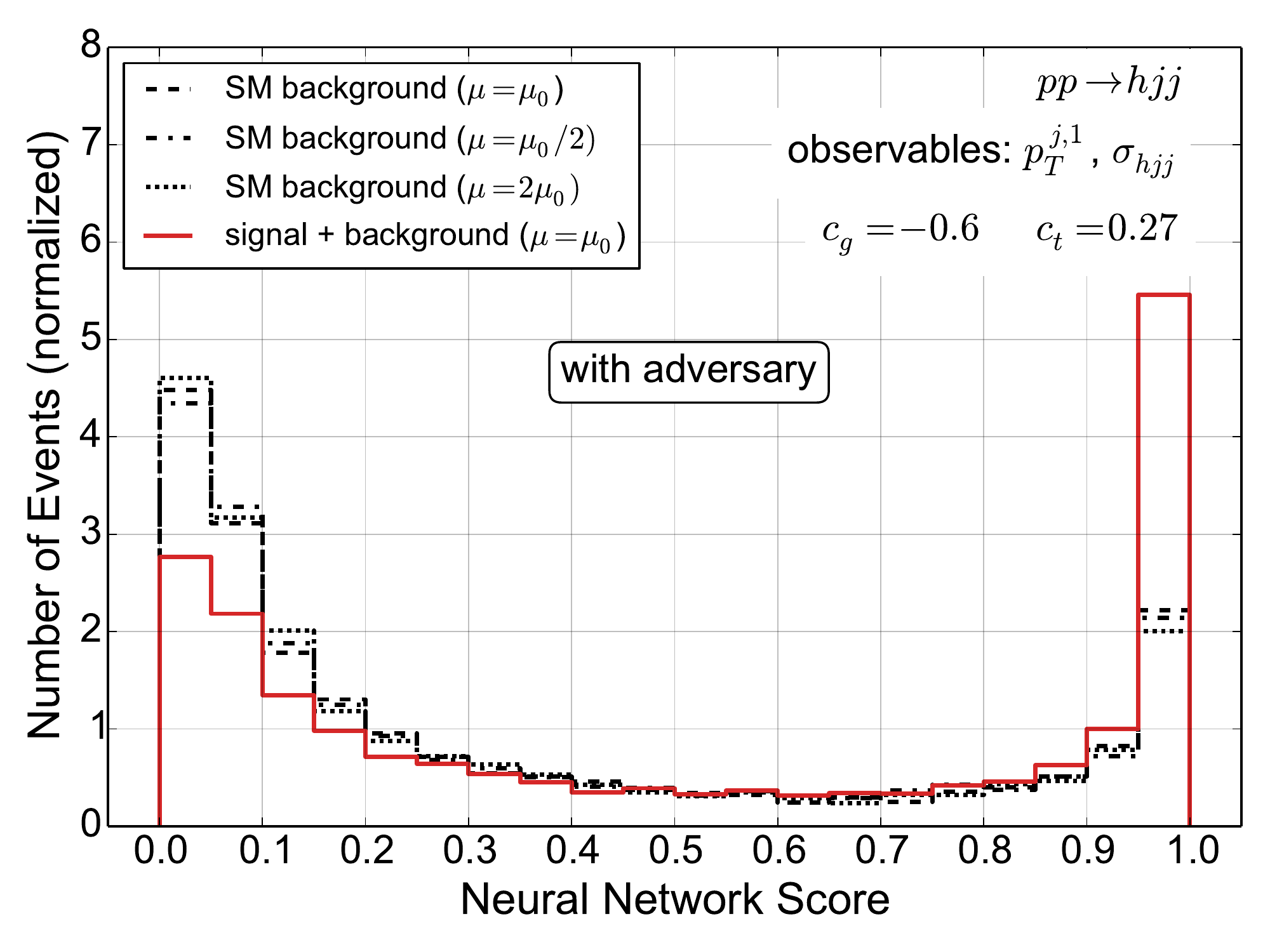}}
\hfill
\subfigure[]{\includegraphics[width=0.48\textwidth]{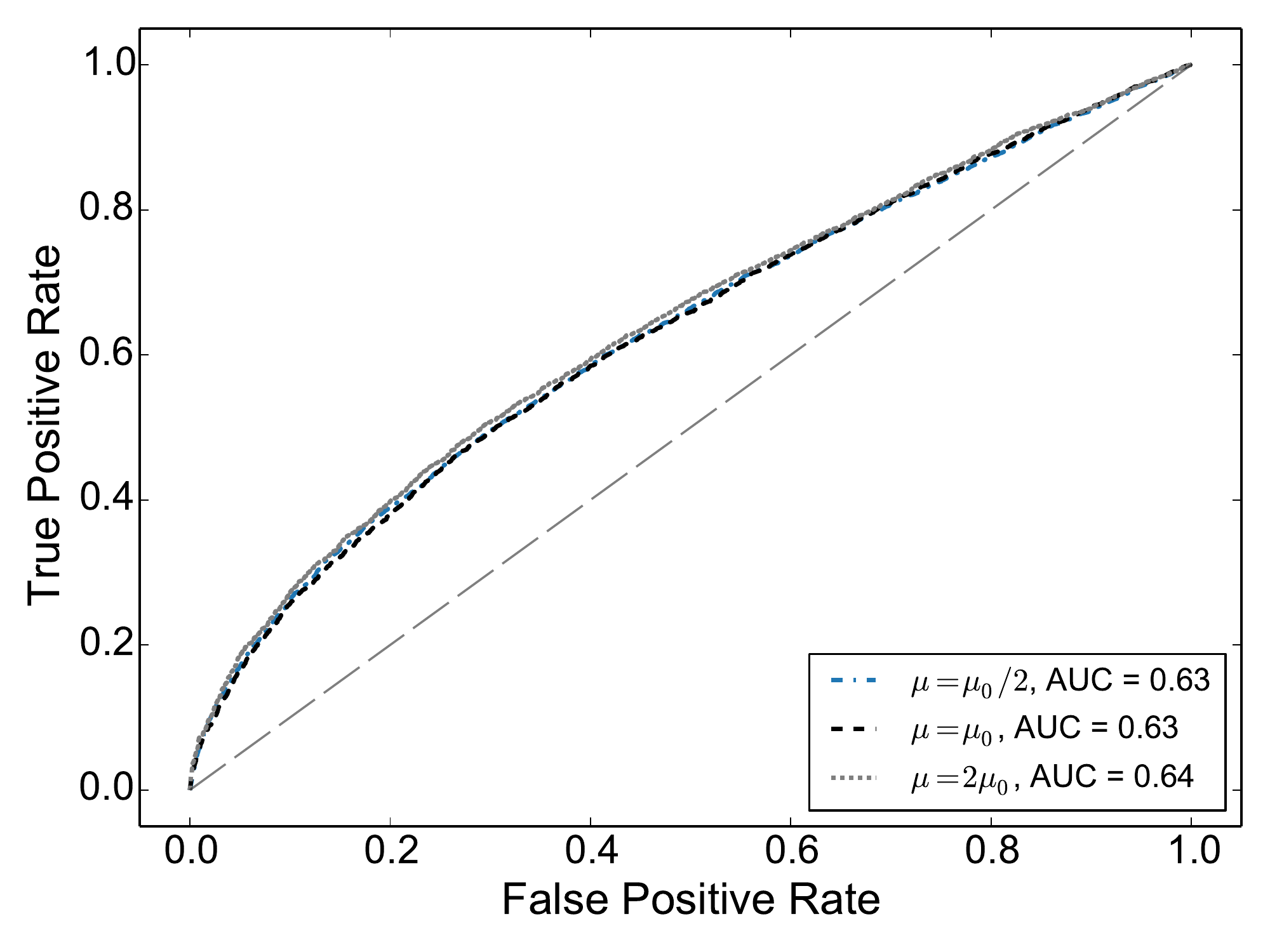}}
\caption{\label{fig:nnexamplescoreadv} Same as Fig.~\ref{fig:nnexamplescorenoadv} but here the distributions were obtained by a classifier that had been trained using the adversarial setup. If the area under curve (AUC) is larger than 0.5, discrimination is possible.}
\end{figure*}

The ANN used here consists of two components. The first component is a classifier discriminating between a standard model Higgs sample and an alternative sample with fixed $c_t$ and $c_g$. The second component is the adversary. This setup is implemented using {\sc{Keras}}~\cite{keras} and {\sc{TensorFlow}}~\cite{Abadi:2016kic}. The classifier has one output node with a softmax activation function, i.e. the output is a scalar $\in[0,1]$ where "0" represents the SM class and "1" the signal class. The classifier output is fed directly into the adversary input. The adversary is trained to determine the scale choice only from the classifier output. Hence, the adversary has one output node with a linear activation function representing the adversary's prediction of the chosen scale.

To perform the adversarial training, we consider a combined loss function consisting of the classifier loss and the adversary loss.
The loss function of the classifier is defined by the binary cross-entropy. The adversarial loss function is defined as a mean squared error regression of the scale. The total loss function is constructed such that the classifier loss contributes positively and the adversarial loss negatively. Hence, the adversarial interplay works as follows: With decreasing ability of the adversary to determine the scale from the classifier output the adversary loss grows. Since it contributes negatively the total loss decreases. The training goal is to minimise the total loss function and therefore the classifier is forced to modify its output such as to minimise the ability of the adversary to distinguish between the scales. This results in a classifier which is insensitive to the scale choice of the input data.

Two architectures exist to perform adversarial training.  An approach where the training of classifier and adversary is performed simultaneously and another with alternating training steps. For the alternating approach the training is also performed on the entire adversarial neural network consisting of classifier and adversary. But in one step the adversary weights are frozen and the total loss function is used. In the other step the classifier weights are frozen and only the adversary loss function is used. Hence, one step trains the classifier taking the adversary penalty into account and the other step trains the adversary only thus adapting the adversary to the previously trained classifier. These two steps are performed alternating on each batch of training data.

We tried both approaches (simultaneous and alternating training), but we found better convergence with the alternating adversary and consequently focused on this approach for this study. For the full NN architecture and training we required: 
\begin{itemize}
\item for the \emph{``classification layer''} 2 hidden layers with 20 nodes each,
\item for the \emph{``adversary layer''} 2 hidden layers with 20 nodes each,
\item in all cases Relu activation function, and 
\item we use a batch size of 500 events trained over 500 epochs. 
\end{itemize}
We have tried other configurations in terms of numbers of layers and nodes but did not observe a significant change in the training performance. However, hyperparameters such as learning rate ($5\times10^{-4}$), relative weight between classifier and adversary loss as well as the number of epochs had to be tuned. To ensure convergence of the adversary, the cross section, jet $p_T$ and any other tested variables are transformed to have mean zero.  The transformation of the cross section is adjusted to have root mean square (RMS) 1, whereas the other variables are transformed to have an RMS of 100. This additional transformation is needed because the scale variation of the adversary and the discrimination power are both dominated by variations in the cross section. To perform the adversarial training the adversary loss is scaled by a factor of 100 relative the loss of the EFT classifier. When the adversary is reduced below 100, for all cases, we observed a gradual transition to the instance where the adversary is non-existent; eventually converging to the bare discrimination case. We use $\sim2.5-4\,\times10^5$ events for signal and background depending on the choice of the parameters $c_g$ and $c_t$. 90\% of the events are used for training and 10\% are reserved for validation and testing.
\subsection{Example}
To highlight the crucial features of our method, we first consider a simple example for which we use our numerical setup given in Sec.~\ref{sec:eft} focusing on the $h$+2~jets channel. For illustration purposes we only consider two input variables in this example: the normalised differential $p_T$ distribution and the associated cross section (see Fig.~\ref{fig:nnexample}). The use of additional variables is studied in Sec.~\ref{sec:secapp}. The choice of $c_g=-0.6,\,c_t=0.27$ is motivated by the shape of the $p_T$ distribution which needs to be contrasted with the overlapping uncertainty bands for the cross sections. We train the NN with background and signal distributions of events defined by the transverse momentum of the leading jet $p_{T,j_1}$ as shown on the right-hand side of Fig.~\ref{fig:nnexample}. The background distributions for all three scale choices in Eq.~\eqref{eq:scales} are combined into one distribution. For the signal we use the central scale ($\mu_0$) distribution. We have checked that the events from $p_{T,j_1}$ distributions of different scales choices produce the same neural network output. The reason is that the NN is only sensitive to shapes since it learns (normalized) probability distributions. However, as can be seen from Fig.~\ref{fig:nnexample} the scale choice has little impact on the shape of the differential cross section with respect to $p_{T,j_1}$.

In addition to the $p_{T,j_1}$ we consider the exclusive $h+2$ jets cross section. We randomly assign to each background (signal) event a cross section distributed according to the background (signal) distribution shown on the left-hand side of Fig.~\ref{fig:nnexample}. Since the theoretical uncertainty of the cross section is estimated by scale variations, its distribution is not governed by statistics. Instead we have to choose a prior. Here we choose an asymmetric Gaussian distribution with mean $\hat{\sigma}=\sigma(\mu_0)$ and left (right) standard deviation $\Delta\sigma_l=\sigma(\mu_0)-\sigma(2\mu_0)$ ($\Delta\sigma_r=\sigma(\mu_0/2)-\sigma(\mu_0)$) to account for the asymmetric character of the theoretical uncertainty associated with the scale choice. We also have checked a flat distribution as a prior and found no significant changes in the NN output of the pivotal classifier as long as the distributions for signal and background cross section overlap.

This is the crucial step in our approach to include \textit{theoretical} uncertainties into the machine learning driven event classification. The key difference to existing approaches to include systematic effects is that in this case the uncertainties affect the event sample as a whole and not event by event, as for example event reconstruction uncertainties. While NNs can be sensitive to theoretical uncertainties which change the shape of event distributions they remain blind to flat uncertainties as in the case at hand. Note that this becomes more important if adapted scale choices exist that capture the shape modifications of certain observables, i.e the ideal scenario of RGE-improved fixed order calculations. Therefore, we propose to promote these theoretical uncertainties to parametrised nuisance parameters to make them accessible on an event-by-event level.

We first run this setup without the adversarial NN. The resulting NN score (and the associated receiver operating characteristic, ROC curve) is shown in Fig.~\ref{fig:nnexamplescorenoadv}. As the uncertainties between the new physics and the SM hypotheses are not necessarily completely correlated we show results for $\mu_0$ there. The classification is highly sensitive to the scale within the boundaries of our scan $1/2<\mu/\mu_0< 2$. There are a number of reasons for such a strong correlation with classification. However, the main qualitative feature that drives this discrimination is captured in the running of strong coupling $\alpha_s$. A feature that is particularly pronounced in the $pp \to h jj$ contribution and our main motivation for the use of this example. The larger the chosen dynamical scale, the smaller the cross section and the larger  the damping of the high $p_T$ tail relative to the central SM choice. In contrast, our choice of non-zero $c_g,c_t$ induces an enhancement of the tail. Together this means that it is easier for the classifier to distinguish the $c_g,c_t$ modification from a lower cross section that results from a comparably soft $p_T$ tail. Conversely, a lower scale choice results in the opposite situation, it is now more difficult for the classifier to distinguish the BSM contribution from a larger cross section that results from an enhanced tail $\sim \alpha_s^4 \log^4(p_T/\mu)$. Note that this is already mitigated in our example as we choose a central scale of $\sim p_T$. Therefore, including the cross section of the whole sample as observable is crucial to isolate scale dependencies of limits, as mentioned above.

The strong dependence of the classifier on scale is noteworthy for measuring BSM-like Higgs properties since it leads to an unphysical response. Close to the blind direction a ``wrong'' choice of $\mu$ could therefore be understood as a measurement of non-zero $c_t,c_g$ in a fit. This is the situation that we need to avoid. 

Fig.~\ref{fig:nnexamplescoreadv} demonstrates, that the adversary eliminates the scale dependence completely. The effect of including the adversary preserves the same discrimination across different scale choices. This means that the particular scale choice does not impact the classification into BSM or SM contribution. More concretely, this means that the NN has learned to avoid regions of phase space parametrized by the physical observables where uncertainties are the key factors that drive the classification in the non-adversary scenario. Put simply, the ANN performs BSM vs SM discrimination only where the SM hypothesis can be trusted. The net effect is therefore not only a convergence of the ROC curves to a single line between $2\mu_0$ and $\mu_0/2$, but an overall reduction of the sensitivity, i.e. three ROC curves that indicate a much reduced, yet reliable, discrimination between signal and SM background.  

\subsection{Application to EFT-modified jet-associated Higgs production}
\label{sec:secapp}

Building on the example of the previous section we can now turn to the multi-dimensional problem of Higgs production in association with up to 2 jets. We apply the numerical setup in Sec.~\ref{sec:eft} by generating Les Houches event files~\cite{Boos:2001cv} for a scan in $(c_g, c_t)$ under the constraint of reproducing the SM-like inclusive cross section within 25\%. Here we consider both Higgs production channels $h$+jet and $h+2$~jets. Furthermore, we treat the cross section for both processes analogously to the example above and additionally employ a range of kinematic information to the classification: for the $h$+jet channel we use transverse momentum and rapidity of the jet and for the $h+2$~jets channel we use transverse momentum and rapidity of the $p_T$-leading and second-leading jet, azimuthal angle between the jets and rapidity and invariant mass of the jet pair. As the uncertainties become limiting factors in particular in the vicinity of the blind direction $c_g-\sqrt{2}c_t/3$, we express the final score as a function of the deviation away from $c_g=\sqrt{2}c_t/3$. 

\begin{figure}[!t]
\includegraphics[width=0.44\textwidth]{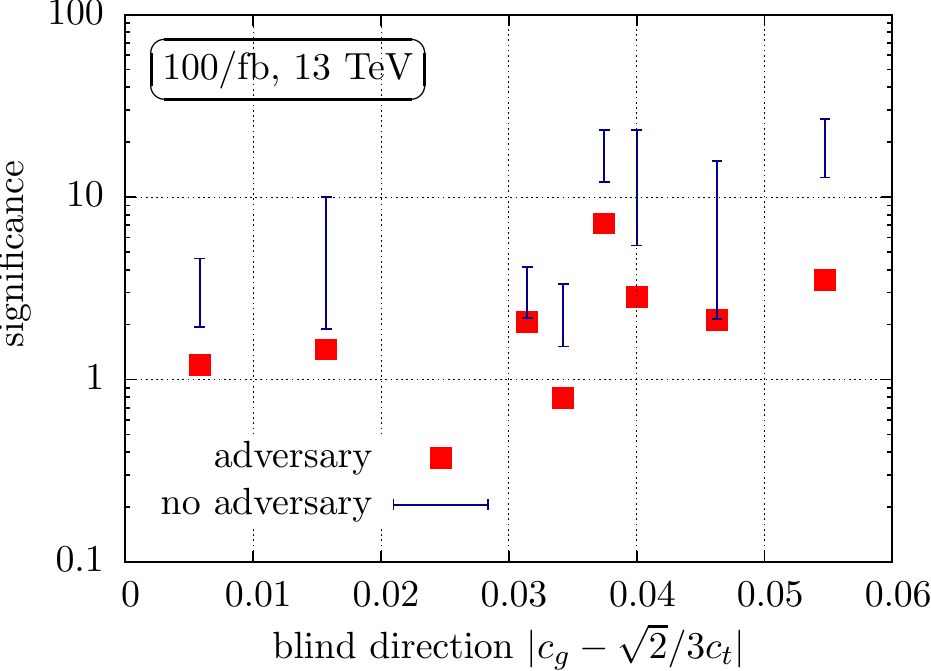}
\caption{\label{fig:ann} Performance comparison of the (A)NN using Higgs+multijet final states. For details see text.}
\end{figure}

The (A)NN output (or ROC curve) can be used to compute significances for different parameter choices. To keep matters transparent, we do this by picking a particular working point on the ROC curve that maximises $S/\sqrt{B}$ (where $B$ stands for the SM expectation), requiring at least 2 (1) expected SM events in the $h+$jet ($h+2$~jets) selection region detailed above for a given luminosity. We treat the two regions as uncorrelated. No additional parton-level cuts are employed.

The result is shown in Fig.~\ref{fig:ann} as a function of the distance from the $(c_g,c_t)$ blind direction for a luminosity of 100/fb. There, we also compare the ANN performance to a neural net analyses without the inclusion of the adversary. In the latter case, different scale choices will result in a different NN score. By tracing the influence of the $\mu$-dependence of the NN score through to the significance, a variation of exclusion can be assigned an uncertainty represented by the blue error bar.

As can be seen from Fig.~\ref{fig:ann}, there are different possible outcomes, but an exclusion at the 68\% confidence level should be possible for the region close to the SM. In some cases the ANN limit agrees well with the lower end of the expected significance as one could naively expect. This situation corresponds to an ANN score that interpolates between maximum and minimum discrimination within the uncertainty bands of the fully differential cross sections. Given that the ANN pivots as a result of the uncertainties, it will always be less sensitive than the NN output. The lower NN sensitivity as a function of $\mu$ therefore provides a supremum of the ANN's sensitivity.

There are also more interesting situations, in particular when we approach the blind direction. While the NN score without adversary becomes sensitive to phase space regions that are not under perturbative control, the ANN will not show any sensitivity in this particular region of phase space. This leads the ANN to push its region of discrimination to a more exclusive region of phase space where the relative impact of the uncertainty is smaller compared to the new physics deviation. In turn, this then manifests itself as a smaller total discriminating power, well outside the naive uncertainty expectation of the NN score without adversary. This robustness is a clear benefit of the adversarial network and is the main result of this analysis. As expected, this effect becomes most relevant when we approach the blind direction. New physics events with $c_g\sim \sqrt{2}c_t /3$ will be distributed more closely to the SM expectation across the considered phase-space. Scale uncertainties render the ANN ``blind'' to small kinematical deviations within the associated uncertainty bands, thereby decreasing the overall sensitivity significantly. Including a proper treatment of kinematic uncertainties, as provided by the ANN is therefore crucial to obtaining robust and reliable constraints that inform a new physics question, which in this example is represented by the relevance of the top threshold for new heavy BSM.

\section{Summary and Conclusions}
\label{sec:conc}
Theoretical and experimental uncertainties are the key limiting factors in searches for new interactions at the LHC and future colliders. This is dramatically highlighted when we want to constrain non-resonant extensions of the Standard Model, where large momentum transfers and very exclusive regions of phase space are the most sensitive probes of new physics. Experimental sensitivities are usually good when we deal with hard final state objects. Unfortunately, outside the inclusive realm of perturbative QCD, theoretical control in highly selective regions of phase space is often lost or at least significantly degraded.

There is no first principle way of correctly assessing the associated theoretical uncertainties apart from ad-hoc scale variations of unphysical remnant scales. Process-dependent QCD-educated guesses for such choices might exist, but these do not come with guarantees, in particular, when we deal with the multi-parton and multi-scale problems imposed by hadron collider phenomenology.

In this paper, we have addressed this conundrum by building on recent developments in machine learning, specifically in the area of adversarial neural networks. 

While ad-hoc scale choices have to remain as estimators of the theoretically unknown, the response of Monte Carlo data to such choices can be propagated to the kinematics of the full final state. In phase space regions where the a priori-sensitivity to new physics is large but effectively obstructed by uncertainties, no sensitivity should be claimed. These regions, which also depend on the particular type of uncertainty, are process-specific and are not necessarily aligned nor connected with our standard understanding of collider kinematics. This large variation in conditions is most naturally addressed with neural networks.

Using the particular case of jet-associated Higgs production at the LHC, where large momentum transfers can pinpoint different sources of new physics in the Higgs sector, we have demonstrated that uncertainties can be accounted for in the discrimination. Additionally we have shown that ``standard'' approaches to select new physics can be sensitive to uncertainties and typically  the sensitivity is over-estimated, in some cases severely. An accurate, uncertainty insensitive estimate, can be achieved through a dedicated adversarial neural network implementation, which provides robust discrimination at expected smaller sensitivity. Although we have focussed on theoretical uncertainties, this methodology directly generalises to other sources of uncertainties that limit the sensitivity of events with high-momentum transfers at the current and future energy frontiers including $b$-tagging efficiencies, jet-substructure calibration, missing energy observables etc. (see in particular Ref.~\cite{Shimmin:2017mfk}) and could be part of a new standard of phenomenological analyses.

\acknowledgements
C.E. is grateful to the Mainz Institute for Theoretical Physics (MITP) for its hospitality and its support during the completion of parts of this work.
C.E. is supported by the IPPP Associateship scheme and by the UK Science and Technology Facilities Council (STFC) under grant ST/P000746/1. 
P.G. is funded by the STFC under grant ST/P000746/1.
P.H. acknowledges the support of the MIT Physics department.

\bibliography{paper.bbl}

\end{document}